\documentclass[review,sort&compress]{elsarticle}

\usepackage{hyperref}
\usepackage{booktabs}

\journal{Nuclear Instruments and Methods in Physics Research Section A}

\begin{document}

\begin{frontmatter}

\title{Improved Plutonium and Americium Photon Branching Ratios from Microcalorimeter Gamma Spectroscopy}


\author[address1]{M. D. Yoho}
\author[address1]{K. E. Koehler\corref{mycorrespondingauthor}}
\author[address3]{D. T. Becker}
\author[address2]{D. A. Bennett}
\author[address1]{M. H. Carpenter}
\author[address1]{M. P. Croce}
\author[address3]{J. D. Gard}
\author[address3]{J. A. B. Mates}
\author[address1]{D. J. Mercer}
\author[address2,address3]{N. J. Ortiz}
\author[address2]{D. R. Schmidt}
\author[address1]{C. M. Smith}
\author[address2]{D. S. Swetz}
\author[address1]{A. D. Tollefson}
\author[address2,address3]{J. N. Ullom}
\author[address2]{L. R. Vale}
\author[address3]{A. L. Wessels}
\author[address1]{D. T. Vo}

\cortext[mycorrespondingauthor]{
		Corresponding author \\
		Tel.: +1 (505) 695 4100 \\
		Email: kkoehler@lanl.gov}

\address[address1]{Los Alamos National Laboratory, Los Alamos, NM 87545, USA}
\address[address2]{NIST Boulder Laboratories, Boulder, CO 80305, USA}
\address[address3]{University of Colorado, Boulder, CO 80309, USA}

\begin{abstract}
Photon branching ratios are critical input data for activities such as nuclear materials protection and accounting because they allow material compositions to be extracted from measurements of gamma-ray intensities.  Uncertainties in these branching ratios are often a limiting source of uncertainty in composition determination.  Here, we use high statistics, high resolution (\~{}60--70 eV full-width-at-half-maximum at 100 keV) gamma-ray spectra acquired using microcalorimeter sensors to substantially reduce the uncertainties for 11 plutonium (${}^{238}$Pu,${}^{239}$Pu,${}^{241}$Pu) and ${}^{241}$Am branching ratios important for material control and accountability and nuclear forensics in the energy range of 125 keV to 208 keV. We show a reduction in uncertainty of over a factor of three for one branching ratio and a factor of 2--3 for four branching ratios.
\end{abstract}


\end{frontmatter}


\section{Introduction}

\paragraph{Microcalorimetry} Recent developments in microwave frequency-division multiplexing \cite{mates2017simultaneous,noroozian2013high,ryhanen1989squid} allow the construction of large superconducting transition-edge sensor (TES) arrays such as the array described in \cite{Becker2019,Bennett2015}. The newly constructed SOFIA (Spectrometer Optimized for Facility Integrated Applications) instrument used in this work currently uses up to 256 pixels with an intrinsic detector efficiency comparable to that of a planar HPGe (high-purity germanium) detector at 100 keV with energy resolutions around 65 eV in the 20--208 keV range \cite{Croce2019_INMM,Yoho2020_umatch}.  

\paragraph{Prior art} The majority of studies of plutonium and \textsuperscript{241}Am branching ratios from 125--208 keV use a set of radionuclide standards (e.g. \textsuperscript{152}Eu or \textsuperscript{166m}Ho) to determine an absolute efficiency curve of an HPGe or Ge(Li) detector \cite{MOREL1994232,helmer1986international,HELMER1985117,willmes1985gamma,yamazaki1966energy,helmer1981emission,HELMER19841067,bortels1984alpha}. With this method, it is necessary to determine the total mass of plutonium, so the masses of purified isotope samples are determined via $\alpha$-spectrometry or isotope dilution mass spectrometry (IDMS). Purified isotopic samples are necessary to reduce the amount of interferences from neighboring signatures. The dominant uncertainty source is from the absolute efficiency curve determination. For example, \cite{helmer1981emission} assigns uncertainties due to calibration radionuclide branching ratios and activity, source absorption, source diameter, and changing detector efficiency over time to determine \textsuperscript{240}Pu branching ratios from 45--160 keV. 

Similar to this work, \cite{baran1994measurement} takes a different approach. Non-isotopically homogenous IRMM (Institute for Reference Materials and Measurements) plutonium standards are counted and plutonium signatures from 125--220 keV inherent in the sample itself are used to determine the relative efficiency curve. In this manner, the branching ratios from 148--161 keV are determined without many of the biases inherent to absolute efficiency determination. This work uses the same approach and fixes five well-known plutonium branching ratios with relative uncertainties ranging from 0.5\% to 1\% taken from the National Nuclear Data Center (NNDC) (see Table~\ref{fixedBrTable}). The excellent resolving power of microcalorimetry, about eight times that of planar HPGe detectors, reduces the systematic biases due to peak interferences, response function fitting, or peak background determination. 

\begin{table}[!htbp]
\caption{Fixed branching ratios from NNDC \cite{ENSDF,ENSDFwebsite}. Uncertainties represent 67\% confidence intervals.\label{fixedBrTable}}
\begin{tabular}{rlrrr}
\toprule
E [keV]	&	Isotope	&	$\gamma$/decay x 100	& Unc. $\gamma$/decay x 100 &	\% Unc.	\\
\midrule
129.30	&	\textsuperscript{239}Pu	&	$6.310 \times 10^{-3}$	& $4.0 \times 10^{-5}$ &	0.63	\\
148.57	&	\textsuperscript{241}Pu		&	$1.8975 \times 10^{-4}$	& $1.25 \times 10^{-6}$ &	0.66	\\
160.31	&	\textsuperscript{240}Pu	&	$4.02 \times 10^{-4}$	& $4 \times 10^{-6}$ &	1.00	\\
195.68	&	\textsuperscript{239}Pu	&	$1.07 \times 10^{-4}$ 	& $1 \times 10^{-6}$ &	0.93	\\
203.55	&	\textsuperscript{239}Pu	&	$5.69 \times 10^{-4}$	& $3 \times 10^{-6}$	& 0.53	\\
\bottomrule
\end{tabular}
\end{table}

\paragraph{Relevance} Hoover, et al. determined that nuclear data uncertainty is the limiting factor in plutonium isotopic analysis~\cite{hoover2013determination}. For example, the prominent \textsuperscript{241}Am $\gamma$-ray peaks at 125.3~keV and 146.7~keV have branching ratio uncertainties of 2.6\% and 2.7\%, respectively. These \textsuperscript{241}Am signatures allow the coupling of the 129.3~keV \textsuperscript{239}Pu $\gamma$-ray to the 104.3~keV \textsuperscript{240}Pu $\gamma$-ray via other \textsuperscript{241}Am $\gamma$-rays below the plutonium K-edge using a relative efficiency curve. The relatively large branching ratio uncertainties on these peaks reduces the accuracy and precision of plutonium material control and accountability measurements which rely on precise measurement of \textsuperscript{240}Pu content. Similarly, the \textsuperscript{241}Am/\textsuperscript{241}Pu chronometer using the strong \textsuperscript{241}Am signature at 146.7~keV and the strong \textsuperscript{241}Pu signature at 148.57~keV is limited by branching ratio uncertainty.     

\section{Experimental}
\begin{figure*}[!htbp]
  \includegraphics[width=1.0\textwidth]{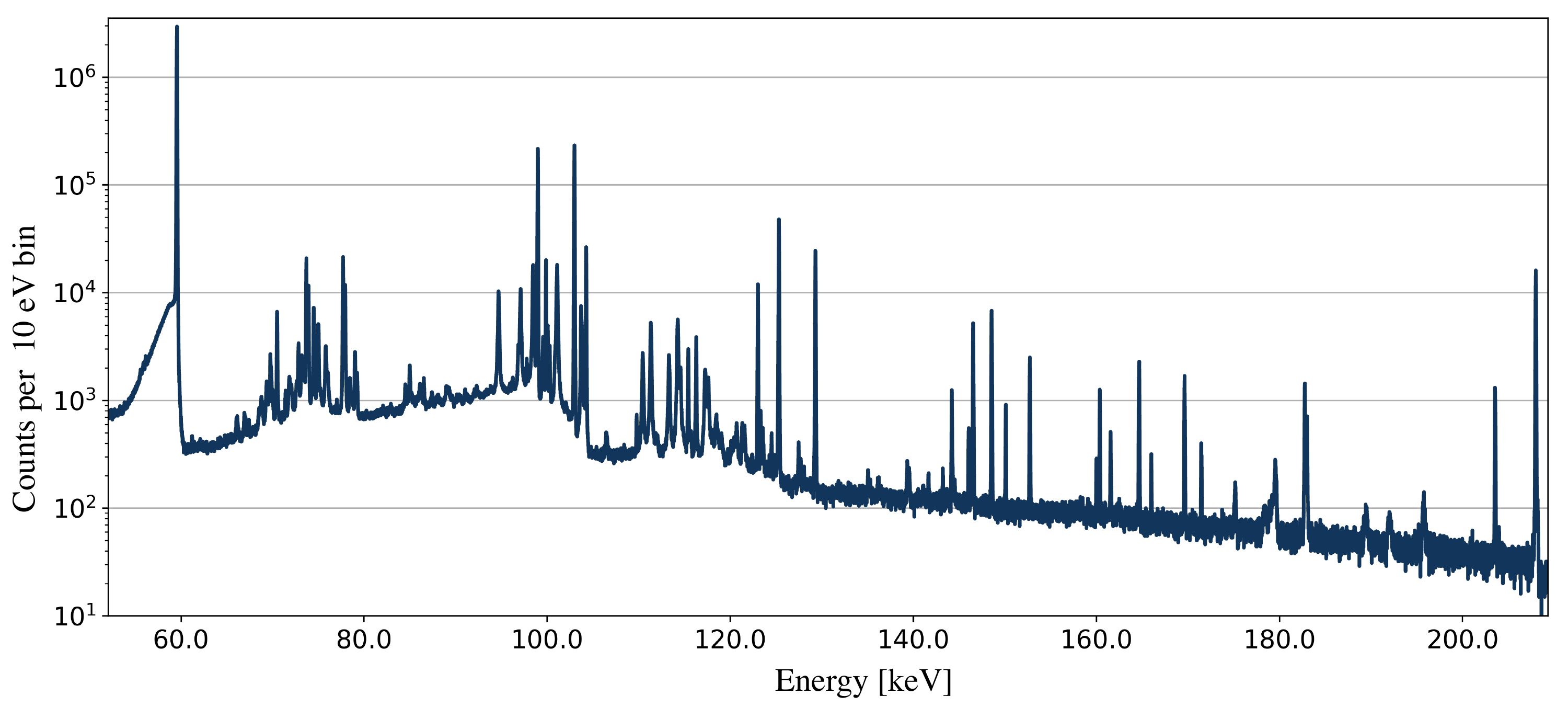}
\caption{CRM137 20 hour spectrum.}
\label{CRM137Spectrum}     
\end{figure*}

\begin{figure*}[!htbp]
  \includegraphics[width=1.0\textwidth]{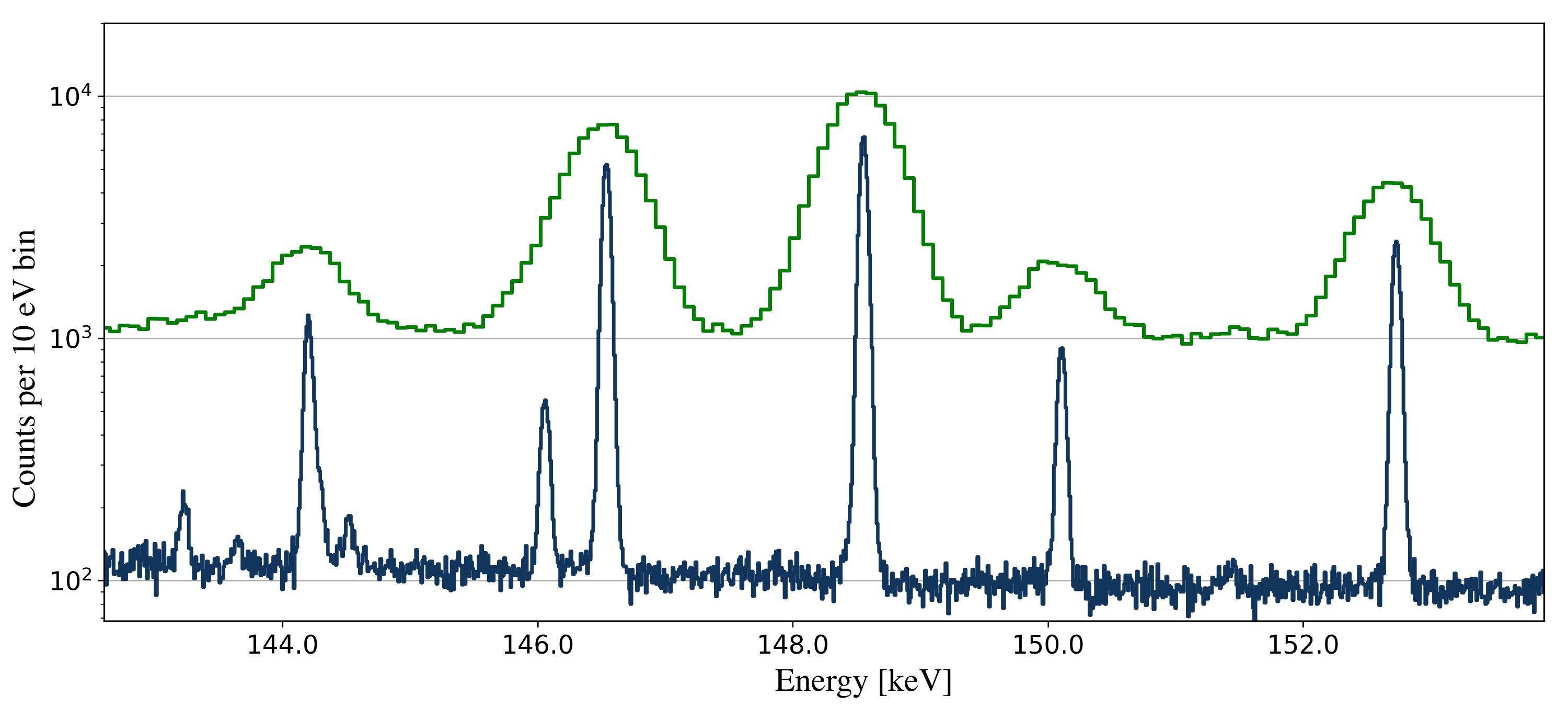}
\caption{150~keV region comparison for 20 hr CRM137 spectrum (bottom) and 1 hr planar HPGe spectrum (top).}
\label{ZoomedIn}     
\end{figure*}

The experimental procedure for acquiring plutonium spectra is well described in \cite{Croce2019_INMM}. Counting conditions and reference materials are described in Table~\ref{IsotopeTable} and Table~\ref{ItemTable}. CRM136, CBNM61, PIDIE1, and PIDIE6 were counted with the BAYMAX (Bimodal Alternate Yield Microcalorimeter Array for X-rays) cryostat using the SLEDGEHAMMER array (Spectrometer to Leverage Extensive Development of Gamma-ray TESs for Huge Arrays using Microwave Multiplexed Enabled Readout) during the period October 2018 to January 2019~\cite{Bennett2015}. All other measurements were made on the SOFIA instrument with the SLEDGEHAMMER array during the period September 2019 to October 2019. One to two mm of Cd filters were used to attenuate the \textsuperscript{241}Am signal at 59.6~keV. Single spectra were acquired for each item except for CRM137, which consists of three separate spectra. Count rates varied from 2 to 12 counts per second per pixel. Figure~\ref{CRM137Spectrum} depicts a typical plutonium spectrum from 60~keV to 208~keV. Figure~\ref{ZoomedIn} demonstrates the excellent resolution of microcalorimetry in comparison to a planar HPGe detector.  

\begin{table}[!htbp]
\caption{Certified and working percent mass fractions with respect to total plutonium. Uncertainties in parentheses represent 67\% confidence intervals. Mass fraction dates are given in Table~\ref{ItemTable}. \label{IsotopeTable}}
\begin{tabular}{llllll}
\toprule
Item	&	\textsuperscript{238}Pu	&	\textsuperscript{239}Pu	&	\textsuperscript{240}Pu	&	\textsuperscript{241}Pu	&	\textsuperscript{241}Am  \\
\midrule
CBNM61	&	1.197(1)	&	62.53(1)	&	25.41(1)	&	6.689(4)	&	1.445(7)	 \\
CBNM70	&	0.8458(9)	&	73.319(5)	&	18.295(4)	&	5.463(2)	&	1.171(6)	\\
CBNM84	&	0.0703(3)	&	84.338(4)	&	14.207(4)	&	1.0275(9)	&	0.217(1)	 \\
CBNM93	&	0.0117(2)	&	93.412(2)	&	6.313(2)	&	0.2235(2)	&	0.105(1)	\\
CRM136	&	0.222(4)	&	84.925(8)	&	12.366(8)	&	1.902(3)	&		\\
CRM137	&	0.267(3)	&	77.55(1)	&	18.79(1)	&	2.168(3)	&		\\
CRM138	&	0.010(1)	&	91.772(5)	&	7.955(5)	&	0.229(1)	&		\\
STDISO3	&	0.006(1)	&	96.302(6)	&	3.562(4)	&	0.111(2)	&	0.0172(4)	 \\
STDISO9	&	0.021(2)	&	92.606(8)	&	6.888(6)	&	0.411(5)	&	0.020(1)	\\
STDISO12	&	0.058(2)	&	86.97(1)	&	11.81(1)	&	0.939(3)	&	0.139(3)	 \\
STDISO15	&	0.169(2)	&	82.11(1)	&	15.41(1)	&	1.604(9)	&	0.068(4)	 \\
PIDIE1	&	0.0111(4)	&	93.765(8)	&	5.990(7)	&	0.199(3)	&	0.228(7)	 \\
PIDIE6	&	0.930(6)	&	66.34(1)	&	23.89(1)	&	5.28(2)	&	3.8(2)	 \\
\bottomrule
\end{tabular}
\end{table}

\begin{table}[!htbp]
\caption{Material size and composition. \label{ItemTable}}
\begin{tabular}{lllll}
\toprule
Item	&	Mass [g]	&	Counts [$\times 10^6$]	&	Count time [h]	&	Certificate date   \\
\midrule
CBNM61	&	6.6 oxide	&	83	&	49	&	20-06-1986	\\
CBNM70	&	6.6 oxide	&	19	&	14	&	20-06-1986	\\
CBNM84	&	6.6 oxide	&	12	&	14	&	20-06-1986	\\
CBNM93	&	6.6 oxide	&	10	&	14	&	20-06-1986	\\
CRM136	&	0.250 sulfate	&	15	&	14	&	01-10-1987	\\
CRM137	&	0.250 sulfate	&	95	&	46	&	01-10-1987	\\
CRM138	&	0.250 sulfate	&	11	&	14	&	01-10-1987	\\
STDISO3	&	11 oxide	&	12	&	14	&	01-07-1986	\\
STDISO9	&	12 oxide	&	15	&	14	&	01-07-1986	\\
STDISO12	&	20 oxide	&	14	&	14	&	01-07-1986	\\
STDISO15	&	12 oxide	&	15	&	20	&	01-07-1986	\\
PIDIE1	&	0.5 oxide	&	164	&	83	&	01-01-1988	\\
PIDIE6	&	0.5 oxide	&	100	&	92	&	01-01-1988	\\
\bottomrule
\end{tabular}
\end{table}

\section{Efficiency Model Validation}
\begin{figure*}[!htbp]
  \includegraphics[width=1.0\textwidth]{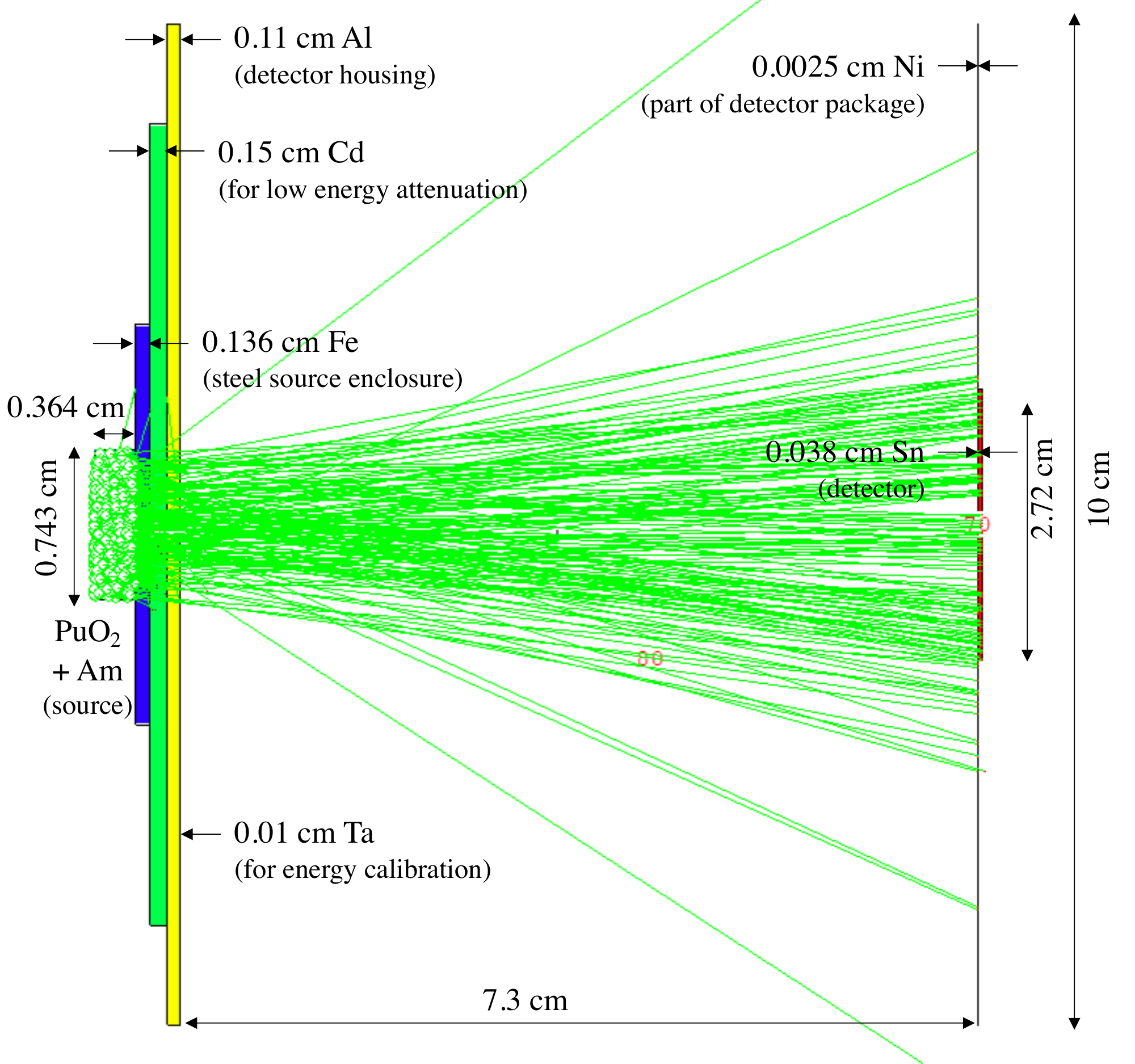}
\caption{MCNP model for CBNM61.}
\label{MCNPModel}     
\end{figure*}

The efficiency model is fit during the optimization routine described in Section~\ref{Algorithm}, but was verified via Monte Carlo modeling with the Monte Carlo N-Particle (MCNP) code version 6.2 \cite{goorley2012initial} for the 0.5~g PIDIE and 5.5~g CBNM reference materials. Figure~\ref{MCNPModel} shows the modeled geometry which consists of the Sn absorbers, PuO\textsubscript{2} with ingrown Am and casings, a Cd attenuator, and the detector package housing. Efficiency curves were then generated by simulating monoenergic photon emissions from 125~keV to 208~keV. The efficiency curve used in this work given in Equation~\ref{effMain} takes into account Sn absorption, Pu attenuation, Cd attenuation, and geometric efficiency. The efficiency $\epsilon_{s}(E(r))$ for spectrum $s$ at energy $E$ associated with region $r$ is described by the physical model

\begin{equation}
\epsilon_{s}(E(r)) = K_{s}g_{s} (1-e^{\mu_{\mathrm{Sn}(E(r))}x_{sSn} } ) e^{-\mu_{\mathrm{Cd}}((E(r))x_{s\mathrm{Cd}}} \frac{ (1 - e ^ {\mu_{\mathrm{PuO}_2}(E(r))x_{s\mathrm{PuO}_2} } ) }{\mu_{\mathrm{PuO}_2}(E(r))x_{s\mathrm{PuO}_2}} .
\label{effMain}
\end{equation}

Here, $K_{s}$ is a scaling factor set for each spectrum such that the efficiency at 129.3~keV is $\simeq 1$. Normalization ensures efficiency values during optimization stay above the machine numerical precision. The terms  $\mu_{\mathrm{Sn}}(E(r))$, $\mu_{\mathrm{Cd}}((E(r))$, and $\mu_{\mathrm{PuO}_2}(E(r))$ represent the Sn photo-electric attenuation coefficients, Cd total attenuation coefficients, and PuO$_2$ total attenuation coefficients at energy $E$ associated with region $r$, respectively. The optimization parameters for each spectrum $s$ denoted by the terms $g_s$, $x_{s\mathrm{Sn}}$, $x_{s\mathrm{Cd}}$, and $x_{s\mathrm{PuO}_2}$ represent geometric efficiency and Sn, Cd, and PuO\textsubscript{2} thicknesses, respectively. Photo-atomic cross-sectional data is taken from the Evaluated Nuclear Data Files (ENDF) B-VIII.0~\cite{cullen} based upon data presented in \cite{cullen1989,cullen2018}. Cubic spline interpolations were used to extrapolate between listed reference energies.

\begin{figure*}[!htbp]
  \includegraphics[width=1.0\textwidth]{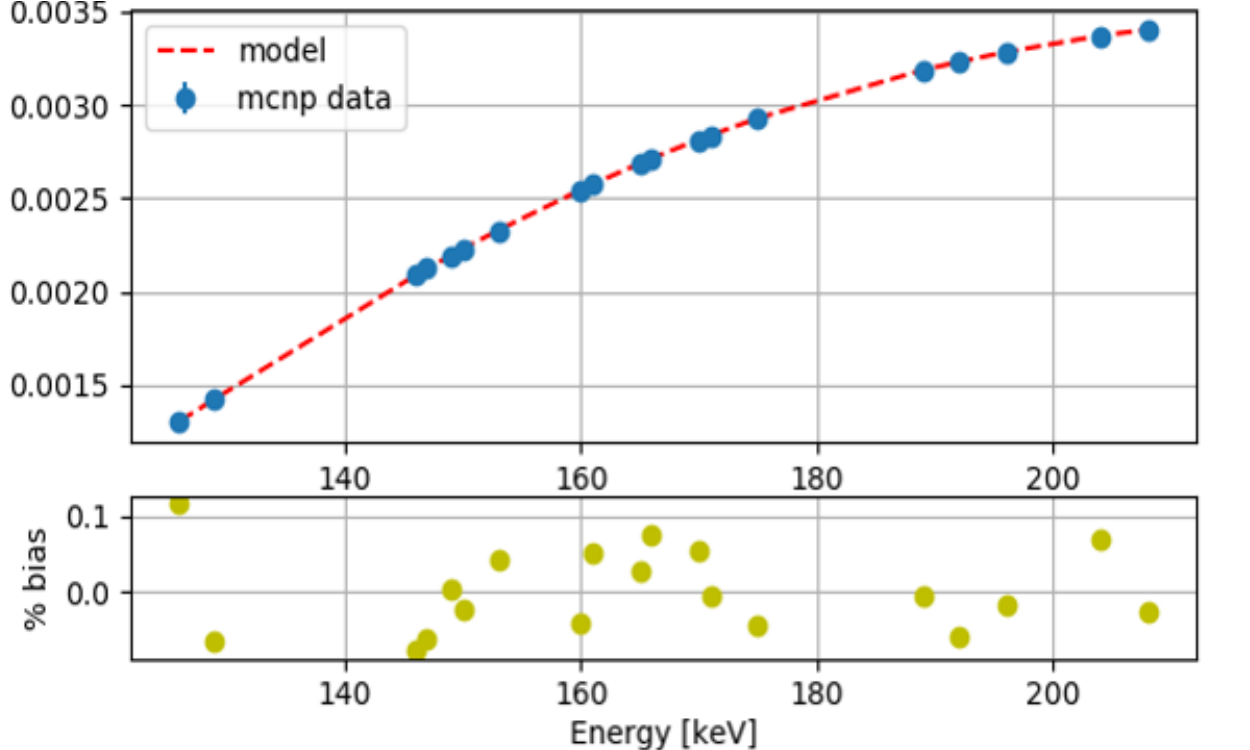}
\caption{Physical efficiency model for CBNM61. The dots are MCNP simulated efficiencies and the line is the fitted physical model. Efficiency units are arbitrary.}
\label{CBNM61SimEff}     
\end{figure*}

A similar physical efficiency model has been used successfully for microcalorimeter data in \cite{Hoover2014} and is very similar to other well-established physical Pu efficiency curves such as in \cite{gunnink1990mga} and \cite{burr2005statistical}. Figure~\ref{CBNM61SimEff} demonstrates that the fit physical efficiency curve describes the simulated data very well with less than 0.1\% bias. Note that there are no visible error bars in the figure since Monte-Carlo simulations were run until there were around $10^7$ full energy deposition events for each simulated energy emission, leading to around a 0.03\% uncertainty for each point which is smaller than the image resolution. The low bias in the fit is due to the fact that efficiency is very gradually curved for the measurement configurations used in this study between 125 and 220 keV. The previous plutonium branching ratio study~\cite{baran1994measurement} over a subset of this energy range reports less than a 0.1\% difference in derived branching ratios using a similar physical efficiency model to Eq.~\ref{effMain} or when using a simple second degree polynomial. This supports the conclusion that many smooth functions will be adequate in this energy range for these measurement configurations. Consequently, \cite{baran1994measurement} does not assign an a priori uncertainty to their chosen second degree polynomial efficiency model between 125 and 220 keV. 

In contrast to this approach, the effect of the choice of efficiency function was explored by repeating the entire analysis described in Sections~\ref{Algorithm} and~\ref{Uncertainty analysis} utilizing a third order polynomial efficiency curve. The derived branching ratios absolutely differed on average by $0.3\sigma$ from those derived using a physical efficiency curve. For this reason, as well as the aforementioned 0.1\% bias on the efficiency curve shown in Figure~\ref{CBNM61SimEff}, a 0.2\% uncertainty component was added in quadrature to the final reported branching ratio uncertainties reported in Section~\ref{Results}.

There are four free parameters in the physical efficiency model, yet the model is fit to five branching ratios (see Table~\ref{fixedBrTable}). The additional branching ratio can be used as a check on the efficiency model fit. In all cases, the fit efficiency curve has a low chi-square.

\section{Algorithm}\label{Algorithm}

\paragraph{Decay correction}
Activity ratios, $\alpha_{is}$, with respect to \textsuperscript{239}Pu for each isotope $i$ for each spectrum $s$, are determined by decay-correcting the mass fractions from Table~\ref{IsotopeTable} to the spectrum measurement dates. For CRMs 136-138, the amount of \textsuperscript{241}Am was taken from the recently published forensics intercomparison exercise analysis of certified reference materials \cite{mathew2019intercomparison}. Decay-corrected activity ratio uncertainties include half-life, mass fraction, and molar mass uncertainties from Table~\ref{isoTable}. The uncertainty for the STDISO series mass fractions was determined by applying the Type~B~On~Bias (BOB) method \cite{levenson2000approach} to the original mass spectrometry reports.  

\begin{table}[!htbp]
\caption{Isotopic data. All uncertainties are 67\% confidence intervals. Half-lives taken from \cite{ENSDF}. \label{isoTable}}
\begin{tabular}{lll}
\toprule
Isotope	&	Half-life [y]	&	Molar mass [g/mol] \\
\midrule
\textsuperscript{238}Pu	&	87.7(1)	    &   238.0495601(19)	\\
\textsuperscript{239}Pu	&	24110(30)	&	239.0521636(19)	\\
\textsuperscript{240}Pu	&	6561(7)	    &   240.0538138(19)	\\
\textsuperscript{241}Pu	&	14.329(29)	&	241.0568517(19)	\\
\textsuperscript{242}Pu	&	3.73(3)$\times 10^5$	&	242.0587428(19)	\\
\textsuperscript{241}Am	&	432.6(6)	&	241.0568293(19)	\\
\bottomrule
\end{tabular}
\end{table}

\begin{figure*}[!htbp]
  \includegraphics[width=1.0\textwidth]{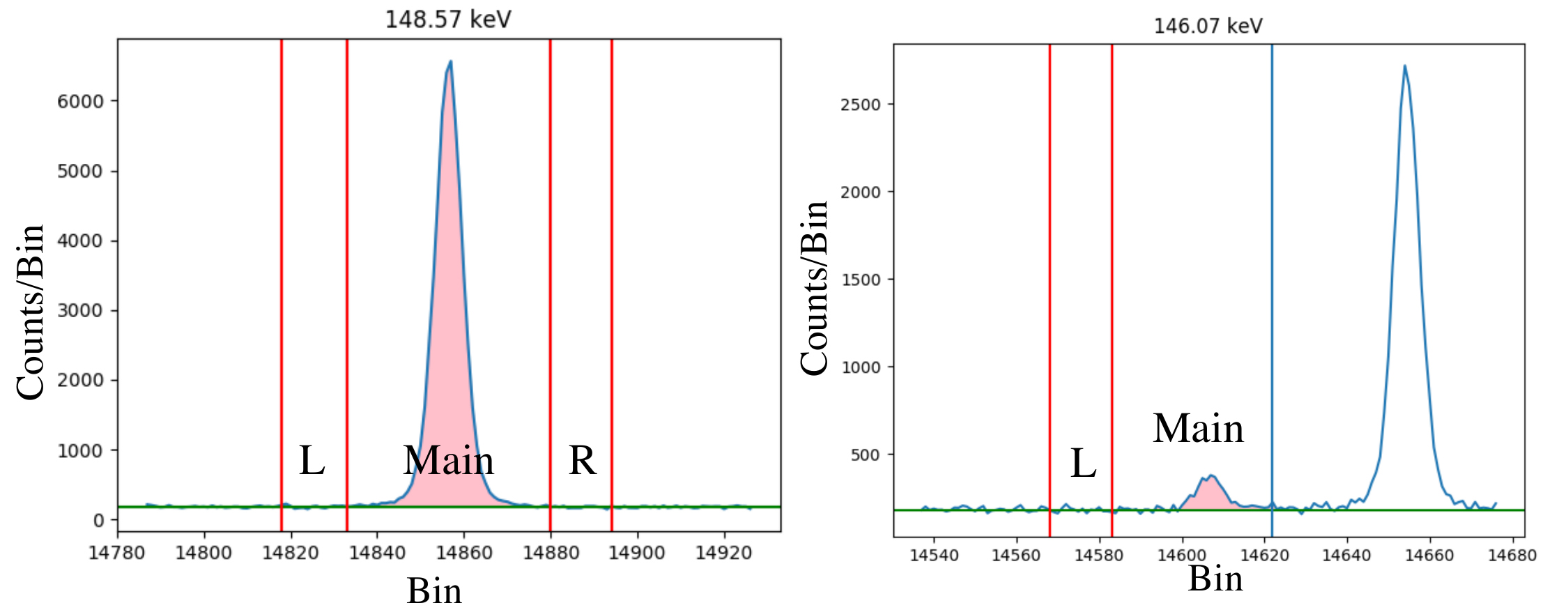}
\caption{Two ROIs delineated for the 148.57~keV peak and 146.07~keV peak. The left and right background regions are demarcated with red lines. The central region labeled ``main'' encompasses at least 99.95\% of the peak area. The 146.07~keV peak has interference on the right side, so the background is calculated using only the left side. (color figure online)}
\label{fig:roi} 
\end{figure*}

\paragraph{Areas}
Net region areas, $A_{sr}$, for each region $r$ associated with each spectrum $s$, are determined via the simple peak integration method described in section 5.4 of~\cite{gilmore2011} and American National Standards Institute (ANSI) N42.14-1999 section C.1~\cite{costrell1999}. This work does not utilize response function fitting, since ANSI standard N42.14-1999 section 6.2~\cite{costrell1999} recommends utilizing the simpler method for isolated singlet peaks. Using this method and assuming a linear background, 
\begin{equation}
A_{sr} = G_{\mathrm{main}} - (G_L + G_{R})C_{\mathrm{main}}/C_{L+R}.
\label{areaEq}
\end{equation}
In Equation~\ref{areaEq}, $G_{\mathrm{main}}$, $G_L$, $G_R$, $C_{\mathrm{main}}$, and $C_{L+R}$ denote the gross counts in a central region, gross counts in left background region, gross counts in the right background region, the number of bins used to calculate $G_{\mathrm{main}}$, and the number of bins used in calculated $G_L$ and $G_R$, respectively. Figure~\ref{fig:roi} shows ROIs for two sample peaks. Left and right background regions generally span five histogram bins. In the case that there is a peak interference on either the left or the right side of the region, that side is assumed to have zero bins and zero counts (e.g. $G_R=0$ and $C_{L+R}=C_L$ as shown in the left panel of Figure~\ref{fig:roi}). Note that in several instances a region spans multiple photon signatures. Central region widths are chosen to encompass \textgreater 99.95\% of the peak area assuming the worst spectrum resolution (75 eV) and a Gaussian peak shape. 

In addition to the Compton background, small-angle Compton scattering will create a step function (sometimes modeled as the error function) underneath each peak. This effect is most pronounced in measurements with thick radiation samples and thick Cd attenuators, which scatter a larger percentage of the photons as they travel toward the detector. This can lead to a bias in measuring large peaks on very small Compton backgrounds. Finally, this effect is more pronounced at higher energies with a larger Compton to photo-electric cross-sectional ratio. The small-angle Compton scattering effect is mitigated by measuring background levels at both lower and higher energies than the primary photo-electric response and averaging the two. In some cases, due to the preponderance of Sn escape peaks, there is no way to estimate the background level on both sides of the peak, so the background level is determined by examining a single side of a peak. 

To estimate the largest possible bias, several large, clean peaks in the largest 10-g CBNM spectra are considered. To estimate the bias in the spectra of low burn-up materials, the net peak area for the ${}^{239}$Pu 203~keV peak from the CBNM-93 spectrum is measured utilizing background regions on the left and right and also measured utilizing only a single background region on the left. This peak is larger in area, higher in energy, and sitting on a smaller Compton background than any peak in the spectrum analyzed with only a single background region. The difference in measured net peak areas is 0.15\%. To estimate the bias in the spectra of high burn-up materials, a similar analysis was conducted on the prominent ${}^{238}$Pu 152~keV peak from CBNM-61 spectrum. For this peak, the difference in measured net peak areas is 0.13\%. This estimated upper bound of a 0.15\% bias is significantly lower than other systematic biases taken into account in this work and is therefore neglected. If the spectra came from larger items that are likely to have high small-angle scattering contributions, such as a kilogram of uranium oxide, then this bias would have to be addressed.

\paragraph{Escape Peaks}
Excited Sn K x-rays have a non-negligible probability of escaping the 0.0380~cm thick absorbers generated by the photo-electric absorption of $\gamma$-rays. These escape peaks interfere with some photon signatures at energies below the primary photo-electric absorption energy. The peaks affected by this effect relevant to this work are given in Table~\ref{EscapeTable}. Other escapes interfering in a region are dealt with by changing the ROI and background bounds.  

Interference-free Sn escapes in multiple high-intensity spectra, such as from the 208~keV \textsuperscript{241}Pu/\textsuperscript{241}Am photo-electric peak, were used to determine the probability of escape (yield) for each x-ray type. Yields for a given Sn escape x-ray (e.g. K$_{\alpha 1}$) were not observed to vary with energy from 125~keV to 208~keV with statistical significance, as verified with Monte Carlo modeling. These yields and uncertainties are given in Table~\ref{EscapeTable}.

\begin{table}[!htbp]
\caption{Sn escape x-rays interfere with relevant peaks for branching ratio determination. These interferences are given below with the primary peak, relevant escape x-ray, and peak being interfered with. The Sn escape x-ray yields are determined from interference-free peaks. Uncertainties in parentheses represent 67\% confidence intervals. The yield is a fraction of the photo-electric peak. \label{EscapeTable}}
\footnotesize
\begin{tabular}{llllllll}
\toprule
\multicolumn{2}{c}{Primary Photo-Peak} & \multicolumn{3}{c}{Sn Escape X-ray} & Escape & \multicolumn{2}{c}{Interference Peak} \\
E [keV]	&	Isotope	&	Type & E [keV] & Yield & E [keV] & E [keV]	&	Isotope\\
\midrule
169.6 & \textsuperscript{241}Am & K$_{\alpha 1}$ & 25.27 & 0.0932(12) & 144.3 & 144.2 & \textsuperscript{239}Pu \\
171.4 & \textsuperscript{239}Pu & K$_{\alpha 1}$ & 25.27 & 0.0932(12) & 146.1 & 146.1 & \textsuperscript{239}Pu \\
175.1 & \textsuperscript{241}Am & K$_{\beta 1}$+K$_{\beta 3}$ & 28.46 & 0.0327(24) & 146.6 & 146.6 & \textsuperscript{241}Am \\
175.1 & \textsuperscript{241}Am & K$_{\alpha 2}$ & 25.04 &	0.0492(14) & 150.1 & 150.0 & \textsuperscript{241}Am \\
188.2 &	\textsuperscript{239}Pu &	K$\beta_1$+K$\beta_3$	 &	28.46 &		0.0327(24)	 &	159.8 &		160.0 &		\textsuperscript{239}Pu \\
189.4 & \textsuperscript{239}Pu & K$_{\beta 2}$ & 29.11 & 0.0068(8) & 160.94 & 160.3 & \textsuperscript{240}Pu \\
203.6 & \textsuperscript{239}Pu & K$_{\beta 1}$+K$_{\beta 3}$ & 28.46 & 0.0327(24) & 174.4 & 175.1 & \textsuperscript{241}Am \\
\bottomrule
\end{tabular}
\end{table}

From the highest energy region areas to the lowest, corrections are made to each $A_{sr}$ with yield $y_v$ associated with escape emanating from region $v$ using

\begin{equation}
A_{sr} := A_{sr} - y_v A_{sv}.
\label{areaUpdate}
\end{equation}

\paragraph{Optimization}
The branching ratio optimization in this work assumes uncorrelated, normally distributed region areas. Therefore, this work uses a $\chi^2$ maximum-likelihood estimator given by

\begin{equation}
\chi^2 = \sum_{s=1}^{N_s} \sum_{r=1}^{N_r} w_{sr} (A_{sr} - \sum_{i=1}^{N_ir}\alpha_{is}\beta_{ir}\epsilon_{s}(E(r)) ).
\label{chisq}
\end{equation}

The weighted differences between the measured net region areas $A_{sr}$ and modeled peak responses are summed for all spectra $s$ and all regions $r$. $N_s$, $N_r$, and $N_{ir}$ represent the total number of spectra, total number of regions, and total number of isotopes with responses in each region $r$, respectively. $\beta_{ir}$ is the branching ratio of isotope $i$ in region $r$. Note that each isotope has at most one response in any given region. The efficiency curve used in the optimization is described in Equation~\ref{effMain}. For each of the 15 spectra, four efficiency parameters ($g_s$, $x_{s\mathrm{Sn}}$, $x_{s\mathrm{Cd}}$ and $x_{s\mathrm{PuO}_2}$) are optimized, resulting in 60 optimization parameters. Additionally, 20 branching ratios are optimized. Therefore, there are 80 optimization parameters total. Each spectrum has 21 measured regions resulting in 235 net degrees of freedom. The bounded, limited-memory approximation of the Broyden-Fletcher-Goldfarb-Shanno (L-BFGS-B) optimization algorithm is used for the non-linear $\chi^2$ minimization \cite{Zhu:1997} All optimization parameters are unbounded. Initial conditions for efficiency parameters are set to be $g_{s}=1$, $x_{s\mathrm{Sn}}= 0.0380$ cm given an Sn density of 7.28~g/cm\textsuperscript{3}, $x_{s\mathrm{Cd}}=0.15$ cm given a density of 8.7~g/cm\textsuperscript{3}, and $x_{s\mathrm{PuO}_2}=0.3$ cm given a density of 10.5~g/cm\textsuperscript{3}. Initial conditions for the branching ratios $\beta_{ir0}$ are randomly selected from a normal distribution with a mean equal to the current ENSDF values and a relative standard deviation of 5\%. 

Randomization of $\beta_{ir0}$ eliminates any potential bias from choosing a set of initial conditions in the neighborhood of a local minimum spanning ENSDF values. As a quality assurance check, the algorithm was run 200 times, and in all cases the algorithm converged to the same minimum. This demonstrates the algorithm solution is independent of $\beta_{ir0}$ within the specified ranges.

The weights $w_{sr}$ in Equation~\ref{chisq} take into account the net area uncertainty $\sigma_{A_{sr}}$ and isotopic ratio uncertainty  $\sigma_{i_{sr}}$ using the effective variance method \cite{orear1982least}. The propagation of uncertainty due to the statistical fluctuation in the Compton background and photo-electric peak for the area calculations is described in ANSI standard N42.14-1999 section C.11~\cite{costrell1999} which is based upon a discussion in Section 5.4.1 of \cite{gilmore2011}. The present work utilizes this method to calculate the net area uncertainties used in the calculation of $w_{sr}$. In this method, the weight term $\delta_j$ associated with $\chi^2$ minimization for function $f$ and dependent measurement point $y_j$ with uncertainty $\delta_{y_j}$ and independent measurement point $x_j$ with an additional uncertainty $\delta_{x_j}$ is given by
\begin{equation}
\delta_j=\left(\frac{\delta f}{\delta x}\right)^2_j (\delta x_j)^2+(\delta y_j)^2.
\label{equiv}
\end{equation} 

Applying Equation~\ref{equiv} to Equation~\ref{chisq} for both non-zero area and activity ratio uncertainties gives
\begin{equation}
\frac{1}{w_{sr}^2} = \sigma_{A_{sr}}^2 + \sum_{i=1}^{N_{ir}}(\sigma_{\alpha_{is}}\epsilon_{s}(E(r))\beta_{ir})^2.
\label{weights}
\end{equation}

Note that Equation~\ref{weights} requires knowledge of the efficiency and branching ratios. This work sets $\beta_{ir}$ = $\beta_{ir0}$. To reduce computational complexity, efficiency $\epsilon_{s}(E(r))$ for the weights is estimated for each spectrum by fitting a 2nd order polynomial efficiency curve in exponential space (see Equation \ref{weights} in \cite{parmentier2018absolute}) to the five fixed branching ratios without using weights. All other uses of the detector efficiency in the optimization algorithm use the physical efficiency model (Equation \ref{effMain}). The algorithm converges in approximately two minutes for a single thread using \~{}16\% of total processing power for an i7-7700 3.60 GHz quad-core processor.

\section{Uncertainty Analysis}\label{Uncertainty analysis}
\begin{figure*}[!htbp]
  \includegraphics[width=1.0\textwidth]{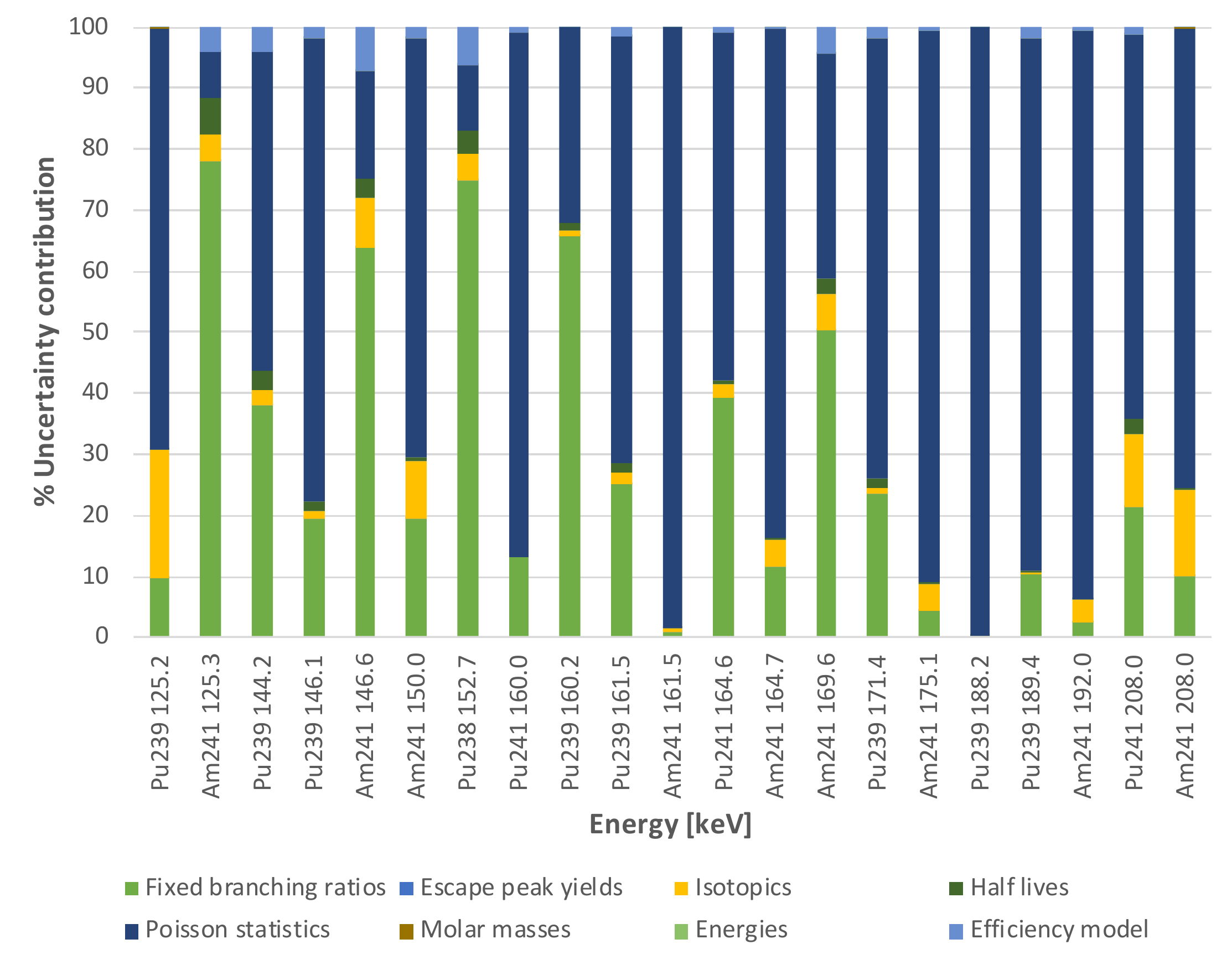}
\caption{Posterior uncertainty budget.}
\label{fig:uncBudget} 
\end{figure*}

This work applies the GUM (Guide to the Expression of Uncertainty in Measurement) Supplement 1 Monte Carlo method \cite{gum2008guide} to determine branching ratio uncertainty. See also \cite{sima2016application} for the application of Supplement 101 to $\gamma$-ray spectrometry efficiency determination. Uncertainty is propagated from Poisson counting statistics, half-lives, molar masses, escape yields, fixed branching ratios, photon energies, and CRM mass fractions. Counts in each  spectrum histogrammed channel are randomly selected from the Poisson distribution. All other parameters are randomly sampled from the normal distribution with a mean equal to the tabulated data or CRM value and standard deviation equal to the tabulated uncertainty. Each Monte Carlo simulation begins prior to CRM mass fraction decay. This captures the correlation between the derived activity ratios for all of the spectra due to the use of the same half-lives. The standard deviations of the optimized branching ratio results for 2000 iterations are taken as the uncertainties. The qualitative uncertainty budget (see Annex~B of Supplement 101 \cite{gum2008guide}) is determined by taking the standard deviation of results from only randomly modulating a single uncertainty component. Due to this qualitative nature, only 200 simulations are run for each uncertainty budget component. The uncertainty budget depicted in Figure~\ref{fig:uncBudget} demonstrates that Poisson statistics uncertainty tends to dominate at higher energies. This is due to the fact that the intrinsic efficiency of the very thin (0.0380~cm thick) Sn absorbers rapidly deteriorates at higher energies.  Some branching ratio uncertainties of important signatures for chronometry, such as those of \textsuperscript{241}Am $\gamma$-rays at 125.3~keV and 146.6~keV are dominated by the uncertainties of the five fixed branching ratios. This uncertainty of around 0.5\% represents an upper-bound that cannot be imporved upon even if higher statistics spectra are acquired.  

\section{Results}\label{Results}

\paragraph{Branching ratios} 
Table~\ref{BRTable} gives the branching ratios and uncertainties determined from this work. All branching ratios agree well within $3\sigma$ of ENSDF values\cite{ENSDF,ENSDFwebsite}. Many branching ratio uncertainties, especially below 160~keV, have been reduced substantially, especially those of \textsuperscript{241}Am. The high branching ratio uncertainties coming from this work (i.e. 125.21~keV, 160.19~keV, and 161.54~keV) are due to insufficient counting statistics and interferences with nearby peaks (i.e. 125.3~keV, 159.955~keV, and 164.45~keV). These branching ratio uncertainties could only be improved with isotope and chemical separations. The high uncertainty for the 188.23~keV peak comes from low statistics and detector efficiency. This could be improved with longer count times. The disagreements between ENSDF values and this work higher than $2\sigma$ primarily come from branching ratios that have high uncertainties. Where uncertainty is substantially improved from ENDSF results (i.e. more than a factor of 1.2), the agreement is in general within $1\sigma$ with the exception of two \textsuperscript{241}Am peaks at 146.55~keV and 150.04~keV. These new values for these peaks with substantially reduced uncertainties are expected to be better than prior values due to using the results of the recently published intercomparison of certified reference materials \cite{mathew2019intercomparison}. The well-measured \textsuperscript{238}Pu branching ratio ($9.46 \times 10^{-4}$ $\gamma$/decay $\times 100$) differs by 1.7$\sigma$ from \cite{ENSDF,ENSDFwebsite}. However, there is excellent agreement with Gunnink \cite{gunnink1976reevaluation} ($9.56 \times 10^{-4}$ $\gamma$/decay $\times 100$)) and Helmer \cite{HELMER19841067} ($9.36 \times 10^{-4}$ $\gamma$/decay $\times 100$)). 

\begin{table}[!htbp]
\caption{Comparison of NNDC \cite{ENSDF} branching ratios (BRs) to those of this work. \textsuperscript{241}Pu BRs at 164~keV and 208~keV assume secular equilibrium with \textsuperscript{237}U. Relative \% uncertainties ($\mu$) represent 67\% confidence intervals. BR units are in $\gamma$/decay x 100. \label{BRTable}}
\footnotesize
\begin{tabular}{lllllll}
\toprule
Energy [keV]	 & 	Isotope	 & 	NNDC BR  & 	$\mu_{BR}$ [\%] 	 & 	This work BR  & 	$\mu_{BR}$ [\%]	 & 	$\mu_{BR}$ Agreement	\\
\midrule
125.21	&	\textsuperscript{239}Pu	&	5.63	$\times 10^{	-5	}$	&	2.7	&	5.51	$\times 10^{	-5	}$	&	13	&	-0.2	\\
125.3	&	\textsuperscript{241}Am	&	4.08	$\times 10^{	-3	}$	&	2.5	&	4.08	$\times 10^{	-3	}$	&	1.0	&	0.0	\\
144.201	&	\textsuperscript{239}Pu	&	2.83	$\times 10^{	-4	}$	&	2.1	&	2.87	$\times 10^{	-4	}$	&	1.0	&	0.6	\\
146.094	&	\textsuperscript{239}Pu	&	1.19	$\times 10^{	-4	}$	&	2.5	&	1.22	$\times 10^{	-4	}$	&	1.4	&	0.7	\\
146.55	&	\textsuperscript{241}Am	&	4.61	$\times 10^{	-4	}$	&	2.6	&	4.75	$\times 10^{	-4	}$	&	0.75	&	1.2	\\
150.04	&	\textsuperscript{241}Am	&	7.40	$\times 10^{	-5	}$	&	3.0	&	7.76	$\times 10^{	-5	}$	&	1.3	&	1.5	\\
152.72	&	\textsuperscript{238}Pu	&	9.29	$\times 10^{	-4	}$	&	0.75	&	9.46	$\times 10^{	-4	}$	&	0.78	&	1.7	\\
159.955	&	\textsuperscript{241}Pu	 &	6.68	$\times 10^{	-6	}$	&	1.1	&	6.87	$\times 10^{	-6	}$	&	2.0	&	1.2	\\
160.19	&	\textsuperscript{239}Pu	&	6.20	$\times 10^{	-6	}$	&	19	&	5.82	$\times 10^{	-7	}$	&	331	&	-2.5	\\
161.45	&	\textsuperscript{239}Pu	&	1.23	$\times 10^{	-4	}$	&	1.6	&	1.20	$\times 10^{	-4	}$	&	1.6	&	-1.1	\\
161.54	&	\textsuperscript{241}Am	&	1.50	$\times 10^{	-6	}$	&	20.0	&	3.52	$\times 10^{	-6	}$	&	19.9	&	2.7	\\
164.61	&	\textsuperscript{241}Pu	 &	4.56	$\times 10^{	-5	}$	&	1.6	&	4.46	$\times 10^{	-5	}$	&	2.0	&	-0.9	\\
164.69	&	\textsuperscript{241}Am	&	6.67	$\times 10^{	-5	}$	&	3.7	&	7.78	$\times 10^{	-5	}$	&	4.9	&	2.4	\\
169.56	&	\textsuperscript{241}Am	&	1.73	$\times 10^{	-4	}$	&	2.3	&	1.72	$\times 10^{	-4	}$	&	0.9	&	-0.3	\\
171.393	&	\textsuperscript{239}Pu	&	1.10	$\times 10^{	-4	}$	&	1.8	&	1.12	$\times 10^{	-4	}$	&	1.4	&	0.9	\\
175.07	&	\textsuperscript{241}Am	&	1.82	$\times 10^{	-5	}$	&	5.5	&	1.85	$\times 10^{	-5	}$	&	2.8	&	0.3	\\
188.23	&	\textsuperscript{239}Pu	&	1.09	$\times 10^{	-5	}$	&	10	&	8.63	$\times 10^{	-6	}$	&	10.8	&	-1.6	\\
189.36	&	\textsuperscript{239}Pu	&	8.30	$\times 10^{	-5	}$	&	1.2	&	7.91	$\times 10^{	-5	}$	&	1.4	&	-2.6	\\
191.96	&	\textsuperscript{241}Am	&	2.16	$\times 10^{	-5	}$	&	4.6	&	2.01	$\times 10^{	-5	}$	&	2.8	&	-1.3	\\
208.005	&	\textsuperscript{241}Pu 	&	5.19	$\times 10^{	-4	}$	&	1.4	&	5.34	$\times 10^{	-4	}$	&	1.9	&	1.2	\\
208.01	&	\textsuperscript{241}Am	&	7.91	$\times 10^{	-4	}$	&	2.4	&	8.08	$\times 10^{	-4	}$	&	5.4	&	0.4	\\
\bottomrule
\end{tabular}
\end{table}

\paragraph{Chronometry} As a quality assurance check, the new measured branching ratios and uncertainties from 125--208~keV were input into a NIST independently developed plutonium isotopic analysis code SAPPY, which is a continuation of work reported in \cite{hoover2013determination} and \cite{Becker2019}. SAPPY uses $\gamma$-ray signatures from 95~keV to 208~keV. The reported \textsuperscript{241}Am/\textsuperscript{241}Pu activity ratios were then used to determine model separation dates \cite{stanley2012beginner} for CRMs 136, 137, and 138. Table~\ref{chronTable} shows improvement in accuracy and precision using the branching ratios derived in this work. Note that CRM documented model ages depicted in Table~\ref{chronTable} are taken from \cite{kirby1984}. CRM136 and CRM137 were well separated via anion-exchange and recrystallization and had no measurable residual \textsuperscript{241}Am. CRM138 was separated via recrystallization and had measurable residual \textsuperscript{241}Am. Therefore, the documented CRM138 separation date in Table~\ref{chronTable} is taken to be the implied purification date on page 4 of \cite{kirby1984}.  

Improvements in accuracy are not significant because the CRM materials were used in the optimization, although the branching ratio optimization uses mass spectrometry isotopic ratios whereas chronometry calculations use documented separation dates as well as an independent analysis code. However, the considerable reduction in uncertainties in separation dates for CRM136 and CRM137 is significant, because it indicates the importance of reduction in branching ratio uncertainty for applications.

The highest statistics spectrum analyzed was a 20 hour measurement of CRM137. Using the branching ratios of this work reduces model age uncertainty by around 50\% from 146 days to 99 days. To show the impact of this uncertainty reduction, this represents a 0.55 \% uncertainty (using a 67\% confidence interval) for a \~{}50 year old reactor-grade plutonium material. 


\begin{table}[!htbp]
\caption{Model age results. Uncertainties represent 67\% confidence intervals. Old denotes the use of NNDC branching ratios. New denotes the use of the branching ratios of this work.\label{chronTable}}
\begin{tabular}{p{3cm}p{2 cm}p{2 cm}p{1.5 cm}p{1.5 cm}}
\toprule
Item	&	Documented separation date \cite{kirby1984}	&	Model separation date & Uncertainty [days] & Difference [days] \\
\midrule
CRM136 (old)	    &	15-Mar-70	&	13-Mar-69	&	183	&	-367	\\
CRM136 (new)	&	15-Mar-70	&	11-Sep-69	&	146	&	-184	\\
\midrule
CRM137 (old)	    &	30-Sep-70	&	22-Apr-70	&	146	&	-157	\\
CRM137 (new)	&	30-Sep-70	&	22-Oct-70	&	99	&	22	    \\
\midrule
CRM138 (old)	    &	12-Jul-62	&	06-Aug-62	&	407	&	25	    \\
CRM 138 (new)	&	12-Jul-62	&	25-Jan-63	&	369	&	197	\\    
\bottomrule
\end{tabular}
\end{table}

\section{Conclusions}
This work has used multiple certified and working reference materials to measure Pu and Am $\gamma$-ray branching ratios from 125--208~keV with microcalorimetry. Many branching ratio uncertainties of decays important for non-destructive plutonium isotopic analysis and nuclear chronometry have been significantly improved. For example, this work reports relative \textsuperscript{241}Am branching ratio uncertainties for $\gamma$-rays at 125.3~keV and 146.65~keV of 1\% and 0.8\% as opposed to the currently listed uncertainties of 2.7\% and 2.6\%, respectively. In an application to the \textsuperscript{241}Am/\textsuperscript{241}Pu parent-daughter ratio for CRMs 136-138 relevant for nondestructive nuclear forensics chronometry, the new branching ratios resulted in improved uncertainty on separation dates. These results support the method of using microcalorimetry for measuring gamma branching ratios. 

The uncertainty budget (see Figure~\ref{fig:uncBudget}) demonstrates that although uncertainty is currently limited by poisson statistics, the ultimate limiting uncertainty comes from the fixed branching ratios which have uncertainties around 0.5\% to 1.0\%.
Future work will entail using improved pixel arrays to get more counting statistics on the well-characterized CBNM and CRM 136--138 reference materials. 

\section{Acknowledgements}
This work was supported by the G. T. Seaborg Institute, the US Department of Energy (DOE) Nuclear Energy's Fuel Cycle Research and Development (FCR\&D), Materials Protection, Accounting and Control Technologies (MPACT) Campaign and Nuclear Energy University Program (NEUP), and the NIST Innovations in Measurement Science program. 

\bibliography{mybibfile}

\end{document}